\documentclass[]{spie}  
\usepackage{amsmath,amsfonts,amssymb}
\usepackage{graphicx}
\usepackage[colorlinks=true, allcolors=blue]{hyperref}
\usepackage{xspace}
\usepackage{subcaption}
\usepackage[export]{adjustbox}

\def\m87{M87$^*$\xspace}
\def\sgra{Sgr~A$^*$\xspace}

\def\lsim{\mathrel{\raise.3ex\hbox{$<$\kern-.75em\lower1ex\hbox{$\sim$}}}}
\def\gsim{\mathrel{\raise.3ex\hbox{$>$\kern-.75em\lower1ex\hbox{$\sim$}}}}

\title{The Black Hole Explorer: Operating a Hybrid Observatory}

\author[1,2,3]{Sara Issaoun}
\author[1]{Kim Alonso}
\author[4,5]{Kazunori Akiyama}
\author[1,3]{Lindy Blackburn}
\author[6]{Don Boroson}
\author[3]{Peter Galison}
\author[1]{Kari Haworth}
\author[1]{Janice Houston}
\author[1,3]{Michael D. Johnson}
\author[7]{Yuri Y. Kovalev}
\author[8]{Peter Kurczynski}
\author[8]{Robert Lafon}
\author{Daniel P. Marrone}
\author[3]{Daniel Palumbo}
\author[8]{Eliad Peretz}
\author[1,3]{Dominic Pesce}
\author[8]{Leonid Petrov}
\author[3]{Alexander Plavin}
\author[6]{Jade Wang}

\affil[1]{Center for Astrophysics $|$ Harvard \& Smithsonian, 60 Garden Street, Cambridge, MA 02138, USA}
\affil[2]{NASA Hubble Fellowship Program, Einstein Fellow}
\affil[3]{Black Hole Initiative at Harvard University, 20 Garden Street, Cambridge, MA 02138, USA}
\affil[4]{MIT Haystack Observatory, Westford, MA 01886}
\affil[5]{Mizusawa VLBI Observatory, NAOJ, Iwate, 023-0861, Japan}
\affil[6]{MIT Lincoln Laboratory, Lexington, MA 02421}
\affil[7]{Max Planck Institute for Radio Astronomy, Bonn, Germany}
\affil[8]{NASA Goddard Space Flight Center, Greenbelt, MD 20771, USA}
\affil[9]{University of Arizona, Steward Observatory, Tucson, AZ 85719, USA}

\authorinfo{Corresponding Author: Sara Issaoun (\url{sara.issaoun@cfa.harvard.edu})}

\begin{document} 
\maketitle

\begin{abstract}
We present a baseline science operations plan for the Black Hole Explorer (BHEX), a space mission concept aiming to confirm the existence of the predicted sharp ``photon ring" resulting from strongly lensed photon trajectories around black holes, as predicted by general relativity, and to measure its size and shape to determine the black hole's spin. BHEX will co-observe with a ground-based very long baseline interferometric (VLBI) array at high-frequency radio wavelengths, providing unprecedented high resolution with the extension to space that will enable photon ring detection and studies of active galactic nuclei. Science operations require a simultaneous coordination between BHEX and a ground array of large and small radio apertures to provide opportunities for surveys and imaging of radio sources, while coordination with a growing network of optical downlink terminals provides the data rates necessary to build sensitivity on long baselines to space. Here we outline the concept of operations for the hybrid observatory, the available observing modes, the observation planning process, and data delivery to achieve the mission goals and meet mission requirements.
\end{abstract}


\section{Introduction}
\label{sec:intro}

\begin{figure}[h]
  \centering
  \centering    
    \includegraphics[width=\textwidth]{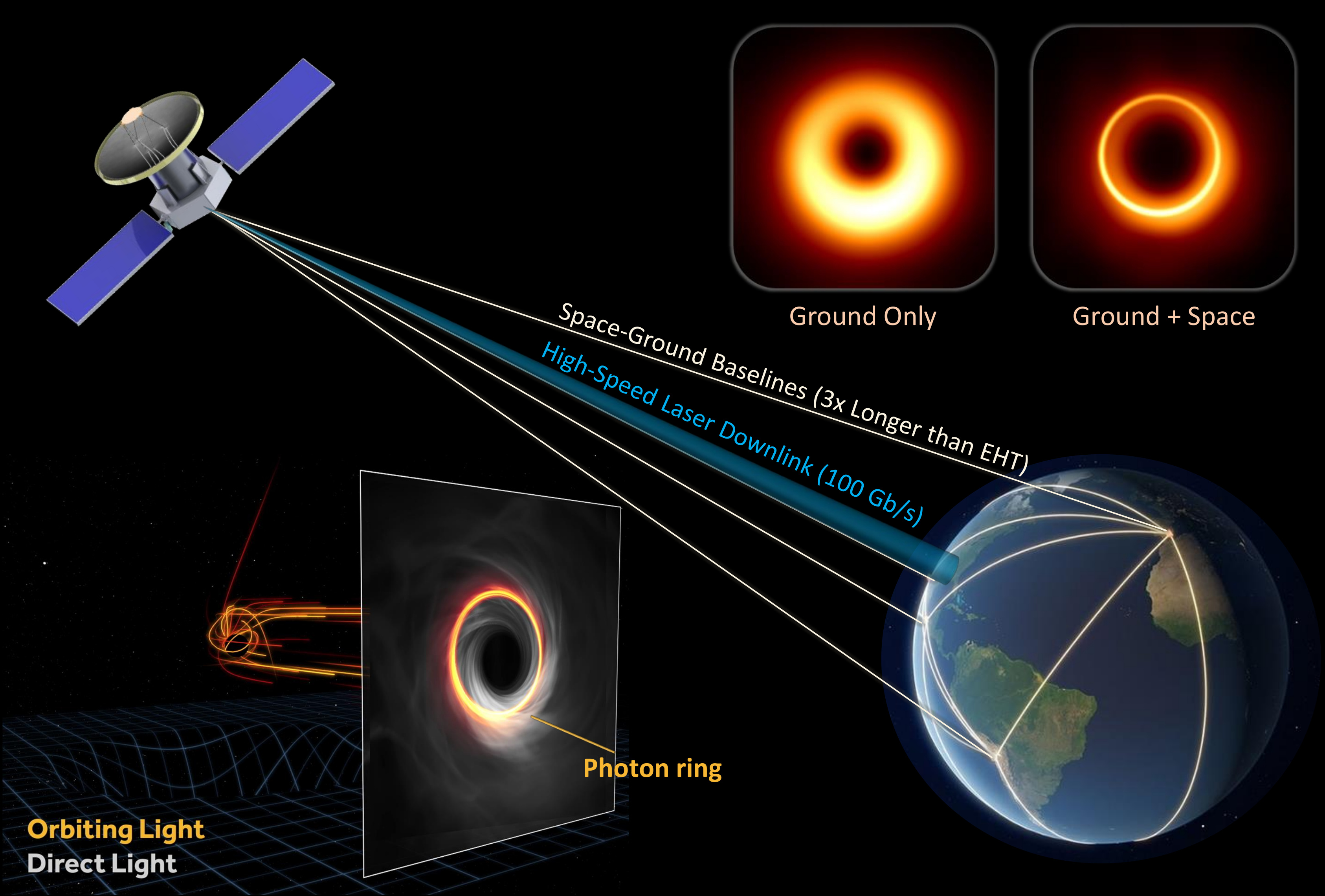}
  \caption{ {\bf The BHEX Mission Concept.} Black hole images display distinctive, universal features such as a sharp ``photon ring'' that is produced from light that has orbited the black hole before escaping. By extending the EHT into space, BHEX will be the first mission to make precise measurements of this striking, untested prediction from general relativity, enabling the first direct measurement of a supermassive black hole's spin. Reproduced from [\citenum{BHEX_Johnson_2024}]. 
  }
  \label{fig:BHEX}
\end{figure}

The Black Hole Explorer (BHEX) mission concept aims to image the sharp ``photon ring" signature around black holes resulting from strongly lensed photon trajectories in extreme gravity, see Figure~\ref{fig:BHEX} \cite{BHEX_Johnson_2024}. Measuring the size and shape of the photon ring would enable a direct measurement of a black hole's spin for two horizon-scale targets: \m87 and \sgra. BHEX will join an existing ground network of radio observatories through very long baseline interferometry (VLBI). With an extension to space, BHEX will provide unprecedented high resolution studies of active galactic nuclei (AGN) jets and a growing population of supermassive black holes in low accretion states.

The BHEX science goals hinge on building sensitivity from high recording bandwidths and optimal atmospheric conditions at the ground very-long-baseline-interferometric (VLBI) stations co-observing with the satellite. Weather considerations also need to be taken into account when developing the downlink network receiving the signal from the satellite. The high bandwidth requirements for BHEX necessitate a downlink infrastructure with optical communications, which have demonstrated higher bandwidth rates than radio-frequency methods commonly used in past space-VLBI experiments. 

\begin{figure}[h]
  \centering
  \centering    
    \includegraphics[width=0.8\textwidth]{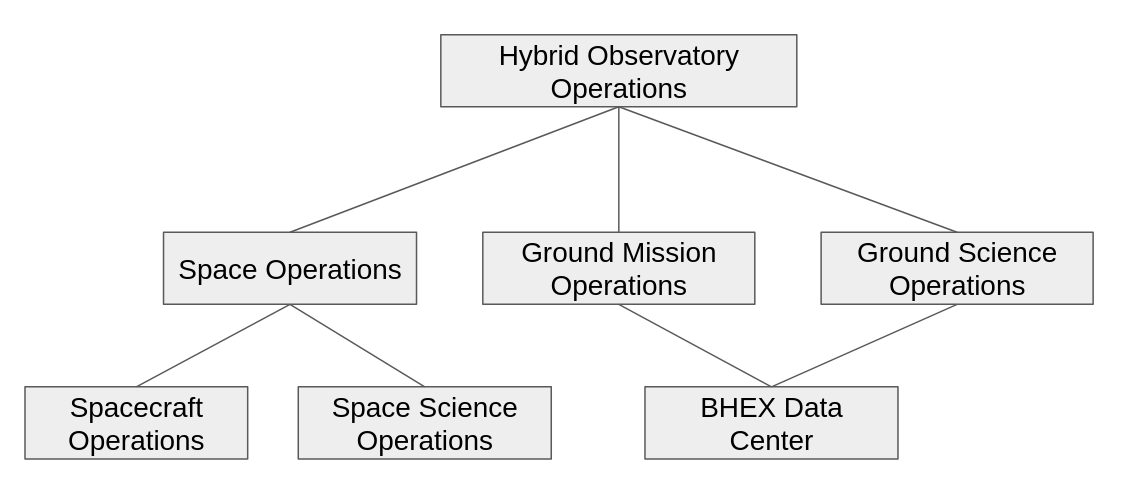}
  \caption{ {\bf Concept of Operations. The BHEX operations involve three parts of a hybrid observatory: the BHEX satellite (with both spacecraft and science operations), the ground VLBI observatories, and the optical downlink network. Data from the ground and space telescopes will be transferred to a data center for correlation, processing, and delivery to a public archive.
  } 
  }
  \label{fig:org_chart}
\end{figure}

The operations concept for BHEX essentially involves a three-part ``hybrid observatory\cite{Mather_2024}'' (see Figure~\ref{fig:org_chart}): satellite operations of the space-based component, coordinated operations of the ground-based VLBI network, and coordinated operations of the ground-based downlink terminals.  

\section{Mission Concept}

BHEX is a space-based component of a space-ground interferometer with a 3.5-meter antenna at medium Earth orbit. The satellite will be equipped with a coherent dual-polarization dual-band receiver system covering 80-106\,GHz and 240-320\,GHz frequency ranges, see Table~\ref{tab:instrument} \cite{BHEX_Marrone_2024,BHEX_Tong_2024}. The receiver system will allow simultaneous dual-band observations, leveraging frequency phase transfer techniques and phase stability at the lower frequency band to build sensitivity at the higher frequency band. The total downlink bandwidth from the satellite will be 64\,Gb/s\cite{BHEX_Srinivasan_2024}. 

\begin{figure}[h]
    \centering
    \includegraphics[width=\linewidth]{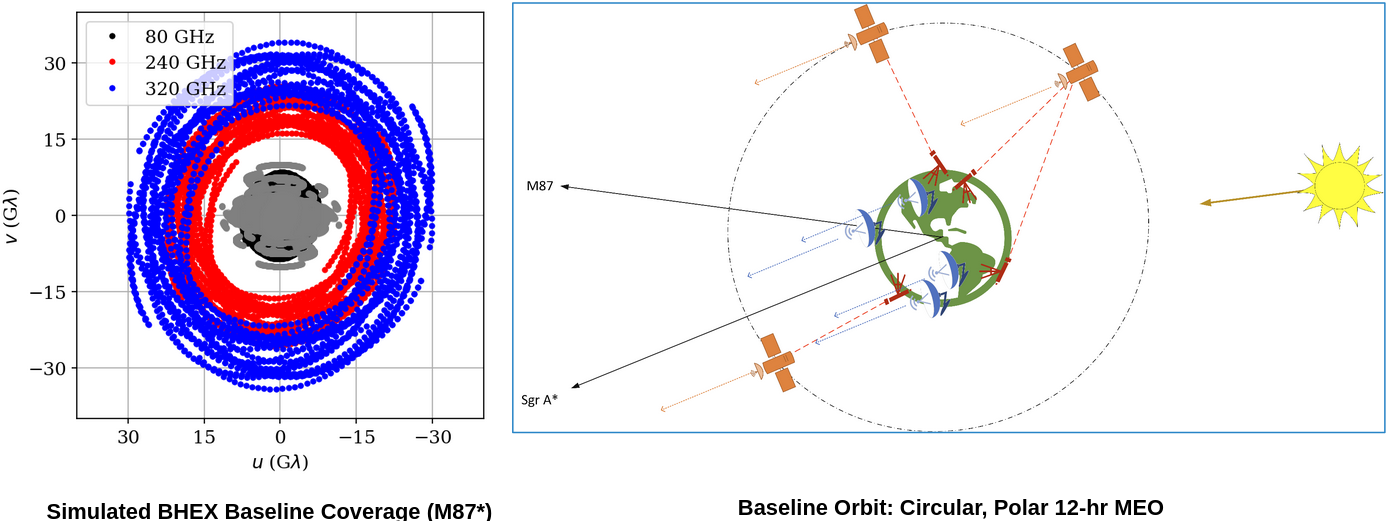}
    \vspace{0.1cm}
    \caption{A nearly polar 12-hr circular orbit will sample the full structure of the photon ring, giving us strong sensitivity to the spin of M87* while producing long N-S coverage necessary for \sgra (because of scattering). Left: Simulated geometrical BHEX coverage on \m87. Right: A circular modestly inclined orbit at MEO. The BHEX satellite is shown in orange, co-observing with the blue VLBI stations and tracked by the red optical downlink terminals. }
    \label{fig:orbit}
\end{figure}

The BHEX orbit will be a 12-hour circular orbit at medium-earth-orbit distances ($\sim$20000 km). This orbit is optimized to obtain circular coverage for \m87, focused on multi-directional sampling of the $n=1$ photon ring signal, while appearing elliptical toward \sgra, focusing on north-south detections of the $n=1$ photon ring where scattering is least affecting \cite{BHEX_Lupsasca_2024}. Figure~\ref{fig:orbit} shows the BHEX ($u,v$) coverage expected for \m87 and a diagram for the orbit. 

\begin{table}[ht]
\caption{BHEX Instrument Specifications.} 
\label{tab:instrument}
\begin{center}       
\begin{tabular}{|l|l|} 
\hline
\rule[-1ex]{0pt}{3.5ex} \bf{Instrument parameter} & \bf{Current specification}  \\
\hline
\rule[-1ex]{0pt}{3.5ex}  Dish diameter & 3.5\,meters   \\
\hline
\rule[-1ex]{0pt}{3.5ex}  Dish surface accuracy & $30-40$\,micron  \\
\hline
\rule[-1ex]{0pt}{3.5ex}  Primary receiver range & $240-320$\,GHz   \\
\hline
\rule[-1ex]{0pt}{3.5ex}  Secondary receiver range & $80-106$\,GHz \\
\hline
\rule[-1ex]{0pt}{3.5ex}  Downlink bandwidth & 64\,Gb/s  \\
\hline 
\hline 
\rule[-1ex]{0pt}{3.5ex}  Ground recording bandwidth & 192\,Gb/s   \\
\hline
\end{tabular}
\end{center}
\end{table}

\subsection{Target Scheduling}

\begin{figure}[h]
    \centering
    \includegraphics[width=0.9\linewidth]{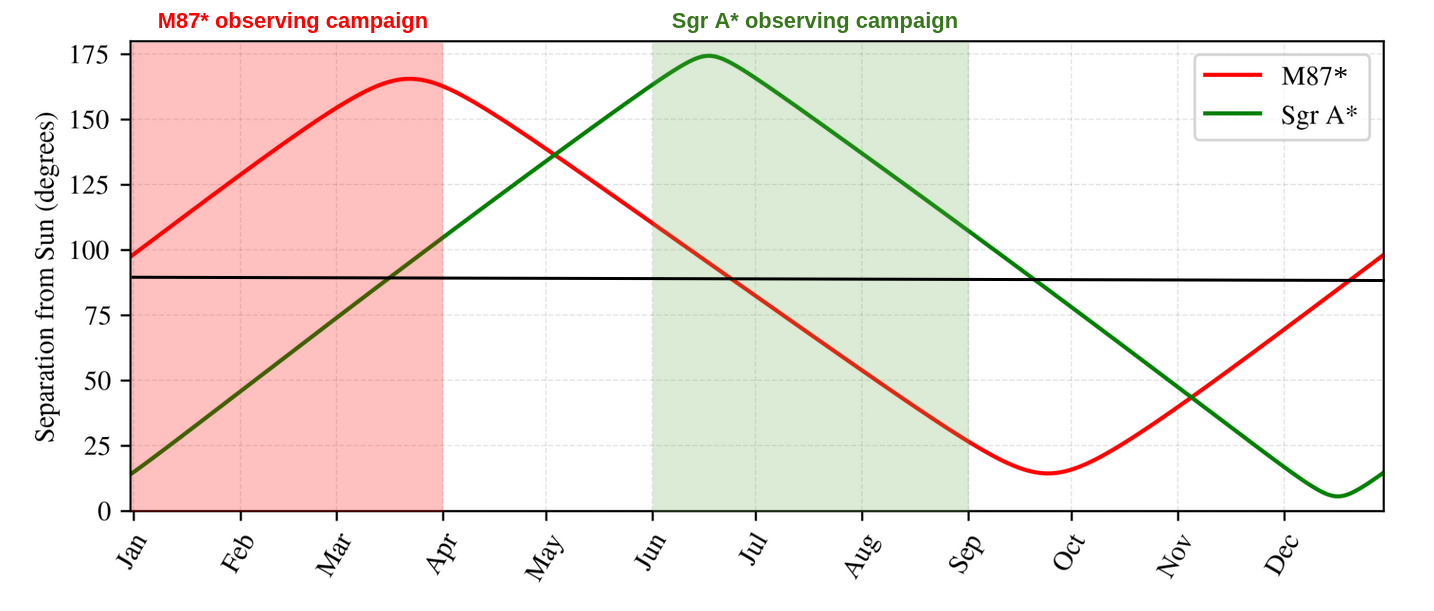}
    \caption{Target separation from the Sun for \m87 and \sgra throughout a calendar year. The three-month observing campaigns are shaded in red and green for \m87 and \sgra, respectively. These campaign ranges correspond to the best overlap of weather trends and night-time observing for the ground VLBI stations.}
    \label{fig:solaravoidance}
\end{figure}

A number of targets are accessible for BHEX, but the participation of the ground VLBI array and sensitivity requirements narrow observing windows around uptime at various sensitive ground observatories and seasonal weather patterns. For the photon ring primary science, solar avoidance (see Figure~\ref{fig:solaravoidance}) and weather considerations for the ground anchor telescopes split observations of \m87 and \sgra to windows in January-March and June-August, respectively.

\begin{figure}[h]
    \centering
    \includegraphics[width=\linewidth]{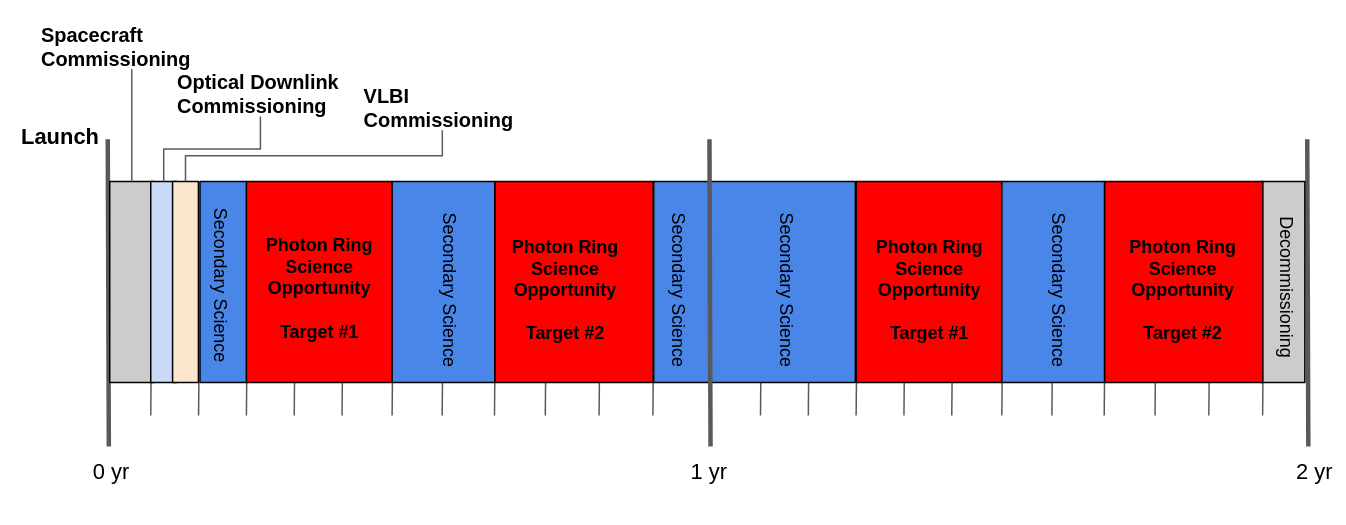}
    \caption{The BHEX mission timeline from launch until decommissioning is a period of two years. Each year will have two key 3-month windows for repeated observations of the photon ring in \sgra and \m87, respectively.  }
    \label{fig:timeline}
\end{figure}

\subsection{Mission Timeline}

The BHEX baseline mission is expected to be operational for a mission length of two years. A timeline is shown in Figure~\ref{fig:timeline}. The first two months after launch will be spent on commissioning of the spacecraft, optical downlink, and VLBI instrument. Each year will have two windows of three months for repeated (at a cadence of every $\sim3$ days) observations of the photon ring in \m87 and \sgra. Time within and outside the windows for the photon ring science is available for additional science i.e., black hole detection surveys, AGN imaging, and single-dish spectroscopy. 

The science program for BHEX will involve high-cadence observations of the main photon ring targets. For \m87, 20 observations will be carried out every three days in a 3-month window aligned with the northern hemisphere winter. For \sgra, 10 observations will be carried out on a best-weather basis in a 3-month window aligned with the southern hemisphere winter. Observations of jet sources and low-accretion black holes can also be scheduled in gaps on the photon ring observing nights depending on their distance to the main target. 

\section{Observatory Operations}

In Figure~\ref{fig:org_chart} we present the concept of operations for BHEX, including both space and ground operations. Below, we summarize the various operations components and their tasks. 

\subsection{Space Operations}

Space operations for BHEX will include two components: spacecraft operations and space science operations. The spacecraft operations include typical spacecraft control, power and thermal management. Spacecraft operations will also maneuver the science instrument, including pointing of the radio telescope antenna and the optical transmission terminal for downlink and telemetry.

Space science operations coordinates observations between the spacecraft and simultaneous ground operations (VLBI observatories, optical terminals). This includes the scheduling of spacecraft observing modes, such as VLBI mode with the ground stations, target-of-opportunity (ToO) science, and space-only ``single-dish'' mode. Space science operations communicates the science and downlink schedules with spacecraft operations. 

\subsection{Ground Operations}

Ground operations for BHEX will include two components: ground mission operations and ground science operations. Ground mission operations coordinates data transfer between the ground and the spacecraft. This includes determining the availability of ground optical terminals for the simultaneous data downlink during science observing, and low-frequency radio stations for Radio Frequency (RF) communications, command \& control, telemetry, and spacecraft tracking. BHEX not have any on-board data storage, so optical downlink must occur simultaneously with observations and coordinated with the space and ground science operations. Availability of the ground optical terminals is determined via weather and cloud cover monitoring, system readiness, and time allocation. Ground mission operations will schedule the downlink and recording of the BHEX science data at the selected optical terminals and coordinate the data e-transfer to the BHEX Data Center. 

Ground science operations are responsible for coordinating VLBI observations at partner radio observatories. Availability of the ground VLBI facilities is determined via weather monitoring, system readiness, and time allocation. Scheduling of the science observations will be done dynamically based on fulfillment of science requirements by the available ground array configuration. Ground operations will schedule the data e-transfer of the ground science data from the VLBI stations to the BHEX Data Center. 

\subsection{Science Observing Modes}

The combination of telescopes forming the ground VLBI array varies with observatory time allocation, weather, and technical capabilities. Based on the BHEX science goals, we can subdivide the ground VLBI participation into two classes:
\begin{itemize}
    \item Imaging Mode: for science goals requiring a substantial imaging capability, primarily for the photon ring science and the jet studies, an extensive ground VLBI array will be coordinated. This will include the participation of at least two partner ground anchor stations (25\,m or larger aperture) and $>4$ smaller dishes to fill the $(u,v)$ coverage for imaging. 
    \item Exploration Mode: this portion of the satellite time will include three partner ground stations, of which two will be of 25\,m or larger diameter to ensure the sensitivity requirements are met. The exploration mode targets visibility domain studies of AGN, transients, and other bright radio sources with flexible scheduling. 
\end{itemize} 

An additional science observing mode, the ``single-dish" mode, will be available for science that does not require a ground VLBI component. This mode will still require coordination of the optical downlink, likely at lower data rates.

\section{Ground Facilities}

The BHEX ground operations involve the simultaneous coordination of both VLBI radio observatories and optical downlink terminals. Suitability of individual sites depends on various factors:
\begin{itemize}
    \item telescope location and mutual visibility with BHEX 
    \item local weather trends (cloud cover, optical depth, atmospheric stability)
    \item telescope sensitivity (aperture size, design, instrumentation)
    \item telescope compatibility (downlink/VLBI equipment, availability) 
\end{itemize}
In the following sections, we discuss the properties of various VLBI and optical stations under consideration for the BHEX ground operations. 

\subsection{VLBI stations}
The sensitivity of continuum VLBI observations depends on the properties of both ends of a particular telescope pair. Assuming a characteristic system-equivalent flux density (SEFD) of $\sim 20,000$\,Jy for BHEX at 240\,GHz, an RMS thermal noise $\sigma_{\rm G-S}$ of $\sim 5$\,mJy is targeted to reach the sensitivity requirement for photon ring detection. The ground-space baseline sensitivity would then be: 
\begin{align}
\sigma_{\rm G-S} &\approx 5\,{\rm mJy} \times 
\left( \frac{{\rm SEFD}_{\rm G}}{1{,}000\,{\rm Jy}} \right)^{1/2} 
\left( \frac{{\rm SEFD}_{\rm S}}{20{,}000\,{\rm Jy}} \right)^{1/2}
\left( \frac{\Delta \nu}{8\,{\rm GHz}} \right)^{-1/2} 
\left( \frac{\Delta t}{100\,{\rm s}} \right)^{-1/2}.
\end{align}
where $\Delta \nu$ is the averaged bandwidth (single polarization), $\Delta t$ is the integration time, and $\eta_{\rm Q} \leq 1$ is a factor that accounts for quantization of the electric field (for BHEX baselines we have $\eta_{\rm Q}=0.75$). 

BHEX will leverage established (sub)millimeter observatories currently observing as part of the EHT. Based on EHT observations, large sensitive apertures, such as the LMT, ALMA, and the IRAM 30-m and NOEMA telescopes, achieve SEFDs on the ground below $1000\,$Jy \cite{EHT2019b,EHT2019c}. Observing with one or two large dishes can successfully anchor BHEX to a wider network of smaller dishes filling the coverage for high-fidelity imaging. 

\begin{figure}[h]
    \centering
    \includegraphics[width=0.49\linewidth]{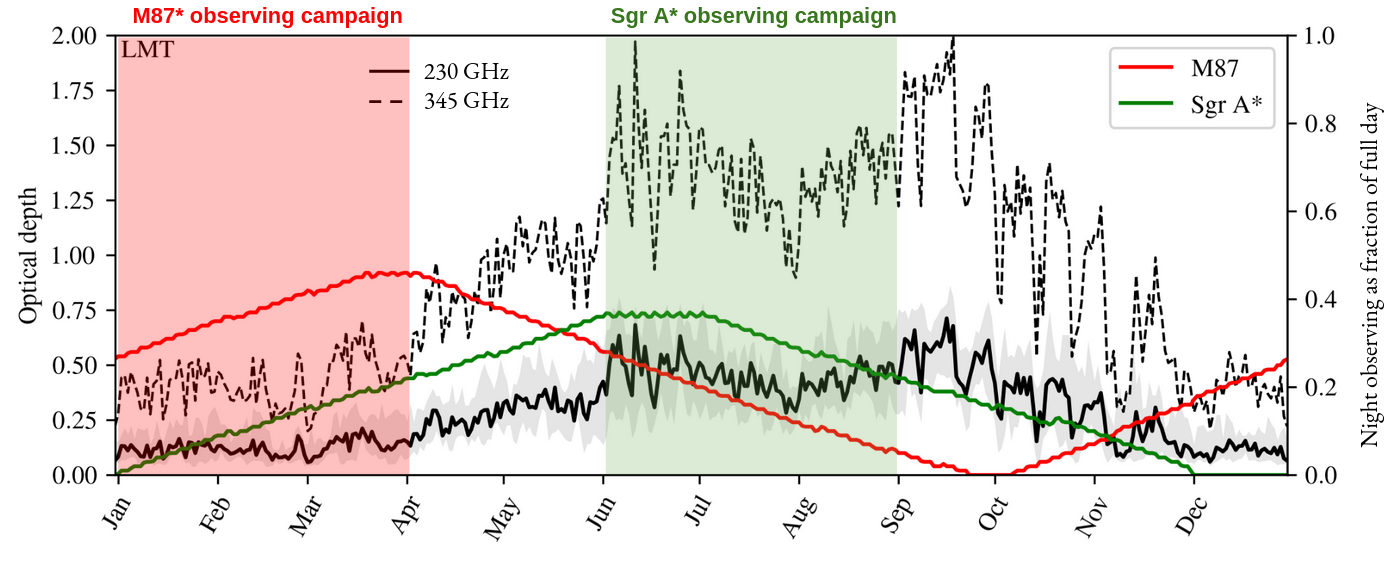}
    \includegraphics[width=0.49\linewidth]{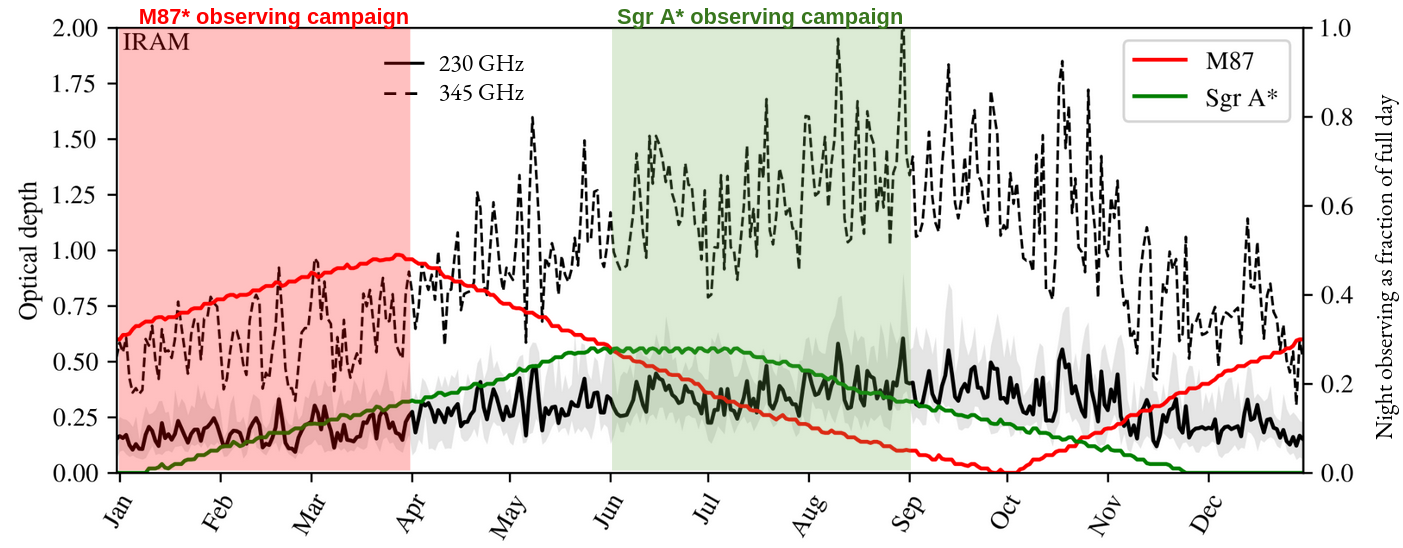}
    \includegraphics[width=0.49\linewidth]{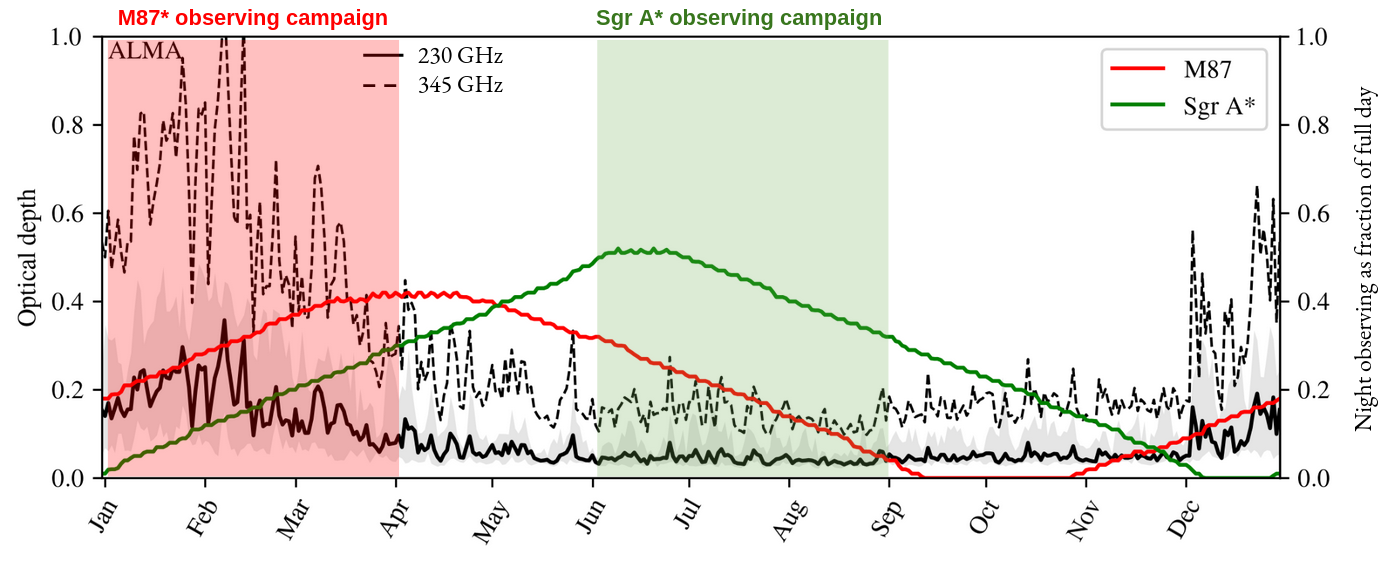}
    \includegraphics[width=0.49\linewidth]{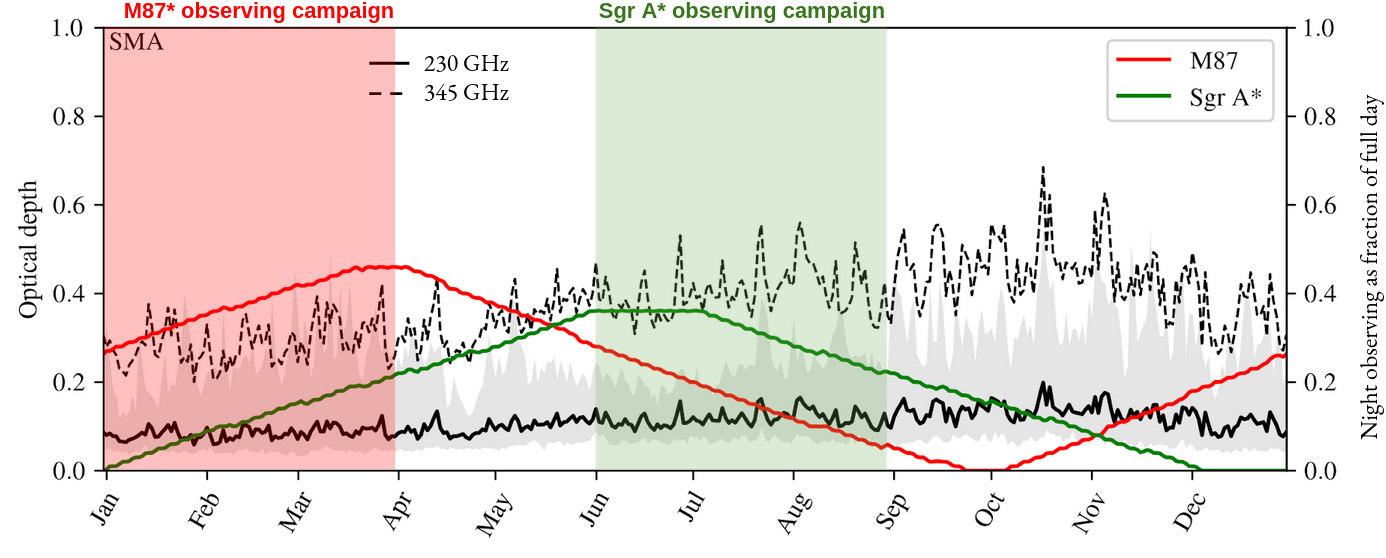}    
    \caption{Optical depth trends over a calendar year at four potential anchor stations, obtained from 44 years of aggregated MERRA-2 data. The optical depth at 230 and 345\,GHz are shown with solid and dashed lines, respectively. The red and green curves represent the uptime at night as a fraction of a full day for \m87 and \sgra, respectively. Based on weather trends, the LMT and IRAM-30m telescope are best suited for the \m87 photon ring campaign, whereas ALMA is best suited for \sgra. The SMA is well suited for both targets. }
    \label{fig:anchor_weather}
\end{figure}

We performed a weather study of potential anchor stations for VLBI science using 44 years of aggregated weather data from the MERRA-2 database. We identified four potential stations that have the receiver range and sensitivity to anchor BHEX to a ground array: the LMT in Mexico, the SMA in Hawaii, the IRAM 30-m telescope in Spain, and ALMA in Chile. All four telescopes operate at the frequencies targeted by BHEX, and are positioned at some of the best (sub)millimeter sites in the world. Of those, we identified ALMA as the best observatory to anchor \sgra photon ring science in the June-August months, and the other three telescopes as most suitable for \m87, see Figure~\ref{fig:anchor_weather}. The SMA is able to support observations of both targets throughout the year due to the high quality of the Mauna Kea site.

\begin{figure}[h]
    \centering
    \includegraphics[width=0.49\linewidth]{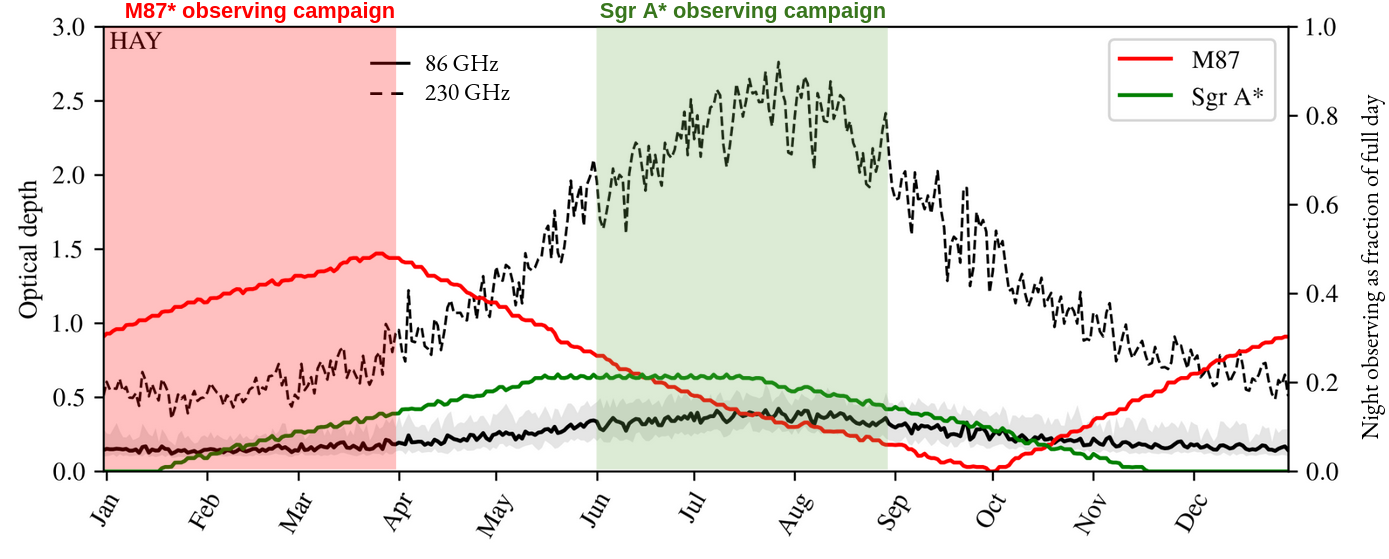}
    \includegraphics[width=0.49\linewidth]{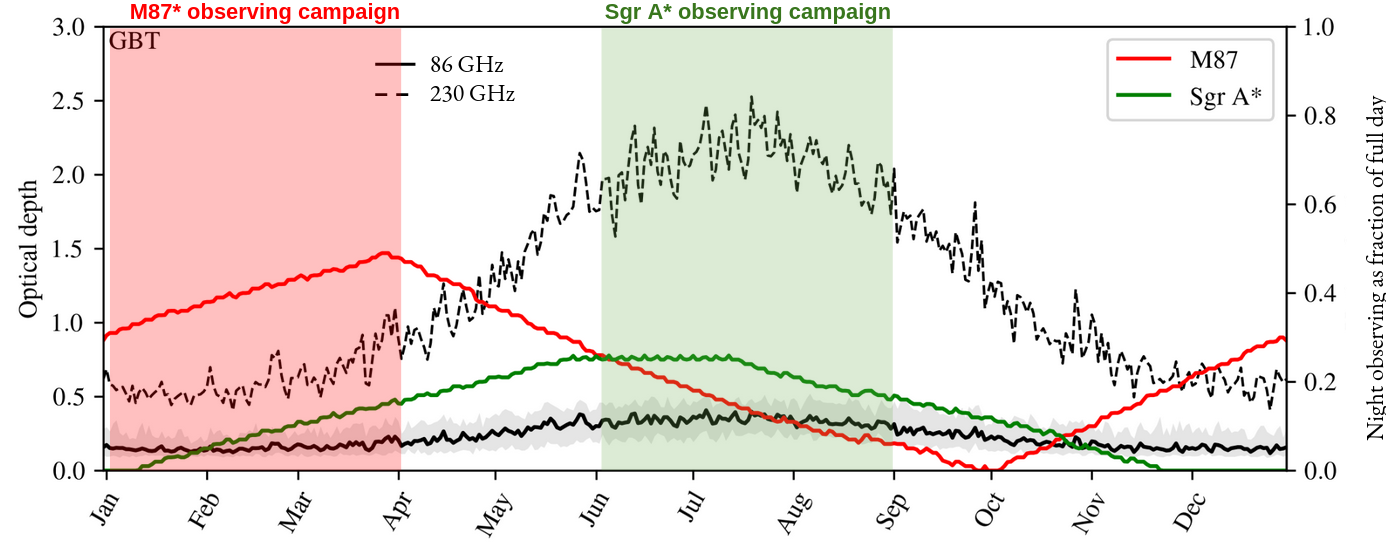}    
    \caption{Optical depth trends over a calendar year at two potential anchor stations for the secondary receiver range, obtained from 44 years of aggregated MERRA-2 data. The optical depth at 86 and 230\,GHz are shown with solid and dashed lines, respectively. The red and green curves represent the uptime at night as a fraction of a full day for \m87 and \sgra, respectively. Based on weather trends, both the Haystack 37-m (left) and the GBT (right) would be able to anchor BHEX at the lower frequency range for the \m87 campaign. }
    \label{fig:86ghz_weather}
\end{figure}

Additionally, we identified the Haystack 37-m telescope and the Greenbank Telescope as potential anchors for BHEX at the secondary frequency (80-100\,GHz). These large sensitive dishes are not at prime sites for higher frequency science, but we can leverage their sensitivity in the secondary receiver range to anchor BHEX to the ground and make use of frequency phase transfer techniques to build sensitivity at the higher frequencies. Both Haystack and GBT would be suitable for \m87 science in the January-March periods, see Figure~\ref{fig:86ghz_weather}. 

\subsection{Optical downlink terminals}

The BHEX downlink system will leverage decades of work to demonstrate laser communication capabilities over a wide range of orbits and missions \cite{Boroson_2014,Caplan_2010,Spellmeyer_2014,Edwards_2018,Khatri_ISS_2023,Khatri_lunar_2023}. Most recently the TeraByte Infrared Delivery (TBIRD) mission demonstrated a downlink rate of 200\,Gb/s from low-earth orbit to ground \cite{Riesing_2023,Schieler_2023}. TBIRD offers an avenue for high data rates with smaller apertures and lower powers that can achieve the data rates needed for BHEX with minimal modifications\cite{BHEX_Wang_2024}. 

Optical frequencies are particularly sensitive to atmospheric scintillation and fading, requiring atmospheric mitigation techniques for error-free data transfer. The selection of ground optical terminals therefore heavily relies on site selection based on atmospheric stability and cloud cover. While optical terminals have been customized for individual experiments, there is a growing need and interest for optical terminals that would be able to service a wide range of missions. 

To guarantee error-free high-bandwidth downlink for BHEX, the selected ground optical terminal network needs to fulfill a number of requirements. These demands were purposed in consideration of space payload and ground terminal capabilities by prioritizing real-time downlink over onboard data buffering, which helps lower power consumption and system mass.
Larger telescope apertures of 0.7 to 3.0 meters can help relax space payload, however in order to preserve signal quality, an adaptive optics (AO) system is needed on the ground terminal. 
In addition, the telescope must support daytime observing to ensure complete downlink coverage and must be compatible with an 8-micron single-mode fiber to couple to the downlink receiver.
Ground station sites must have a favorable weather conditions with minimal cloud coverage to maximize orbital coverage. These considerations guide the selection of ground stations most compatible with the BHEX mission's requirements.

It is ideal to partner with sites that operate exclusively for downlink to eliminate the need for modifications, and down-select within proximity to our radio telescopes for the coordination of data transfer. A number of facilities stand out while assessing possible terminals. 

The optical station Low Cost Optical Terminal (LCOT) managed by NASA Goddard is specifically designed to support laser communication, offering features we seek in our terminal sites for \sgra detection window. The NICT ground terminals consist of four 1-2 meter telescopes to be distributed throughout different locations in Japan, ensuring at least one terminal can operate optimally under varying atmospheric conditions \cite{BHEX_Akiyama_2024}. The Optical Ground Station-2 located in Haleakala, Hawai'i, offers optimal weather conditions, ideal for the photon ring campaigns of both targets. Despite its 0.6 meter aperture, its strategic location and design for laser communications make it an important asset for mission operations.

Additionally, there are several other optical telescope facilities that, although they were not originally designed for laser downlink, have had collaboration in other laser communication projects, therefore they could be modified to meet our requirements, see Table~\ref{tab:terminals}. For instance, the Lowell Observatory in Perth is equipped with a 0.6-meter telescope, the ESO facility in Chile features a 2.2-meter telescope, and the Aristarchos 2.3-meter aperture telescope is located in Greece. These sites hold great potential due to their large apertures and clear skies. Many of these telescopes were also suggested because they are compatible with the observing campaign for our second target, \sgra. The \sgra campaign benefits by having a significantly wider range of location terminals specifically designed for laser communication.

\begin{table}[ht]
\caption{BHEX Potential Downlink Terminals.} 
\label{tab:terminals}
\begin{center}       
\begin{tabular}{|l|l|l|l|} 
\hline
\rule[-1ex]{0pt}{3.5ex} \bf{Station} & \bf{Location} & \bf{Aperture} & \bf{Prime Target}\\\hline
\rule[-1ex]{0pt}{3.5ex}  LCOT & Maryland, USA & 0.7 m & \sgra \\\hline
\rule[-1ex]{0pt}{3.5ex}  NICT & Japan (multiple locations) & 1-2 m & \m87 \\\hline
\rule[-1ex]{0pt}{3.5ex}  ESO & La Silla, Chile & 2.2 m & \m87 \\\hline
\rule[-1ex]{0pt}{3.5ex}  ARIS & Achaea, Greece & 2.3 m & \sgra \\\hline
\rule[-1ex]{0pt}{3.5ex}  LOWELL & Perth, Australia & 0.6 m & \m87 and \sgra \\\hline  
\rule[-1ex]{0pt}{3.5ex}  OGS-2 & Haleakala, Hawaii & 0.6 m & \m87 and \sgra \\\hline
\rule[-1ex]{0pt}{3.5ex}  OGS-Teide & Tenerife, Canary Islands & 1 m & \m87 and \sgra \\\hline
\rule[-1ex]{0pt}{3.5ex}  CAHA & Almeria, Spain & 1-2 m & \sgra \\\hline
\rule[-1ex]{0pt}{3.5ex}  KUIPER & Arizona, USA & 1.5 m & \sgra \\\hline
\end{tabular}
\end{center}
\end{table}

\begin{figure}[h]
    \centering
    \includegraphics[width=0.9\linewidth]{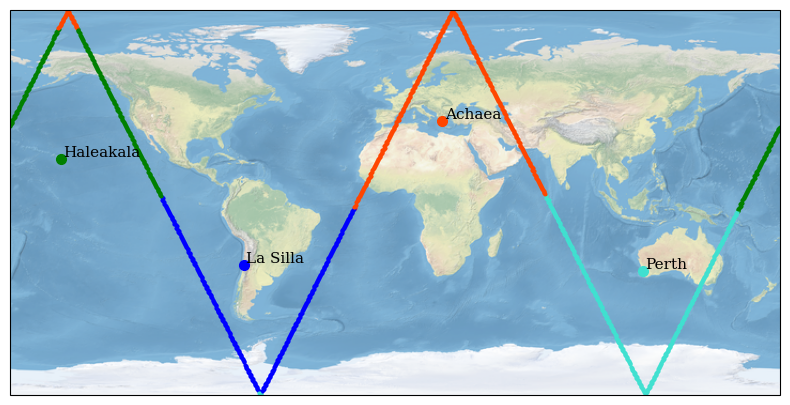} 
    \caption{This map represents the four identified downlink locations on Earth for 24-hour coverage for BHEX observations of \m87, indicated by bullets. The trajectory is color-coded to indicate the coverage fraction of the orbit by each terminal as BHEX passes within 15\,degrees elevation overhead.}
    \label{fig:M87_potential_stations}
\end{figure}

\begin{figure}[h]
  \centering
  \includegraphics[width=0.49\textwidth]{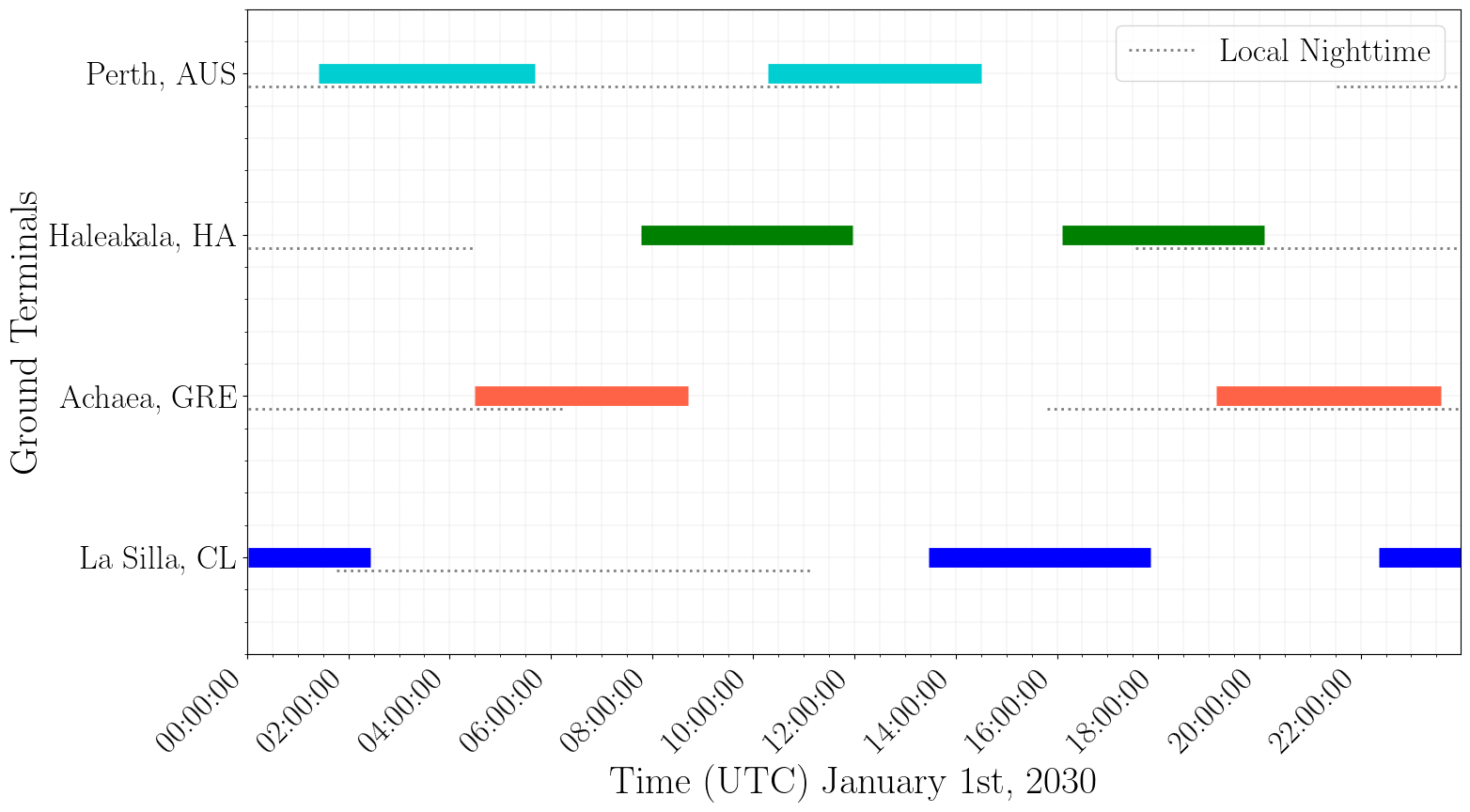} 
  \includegraphics[width=0.49\textwidth]{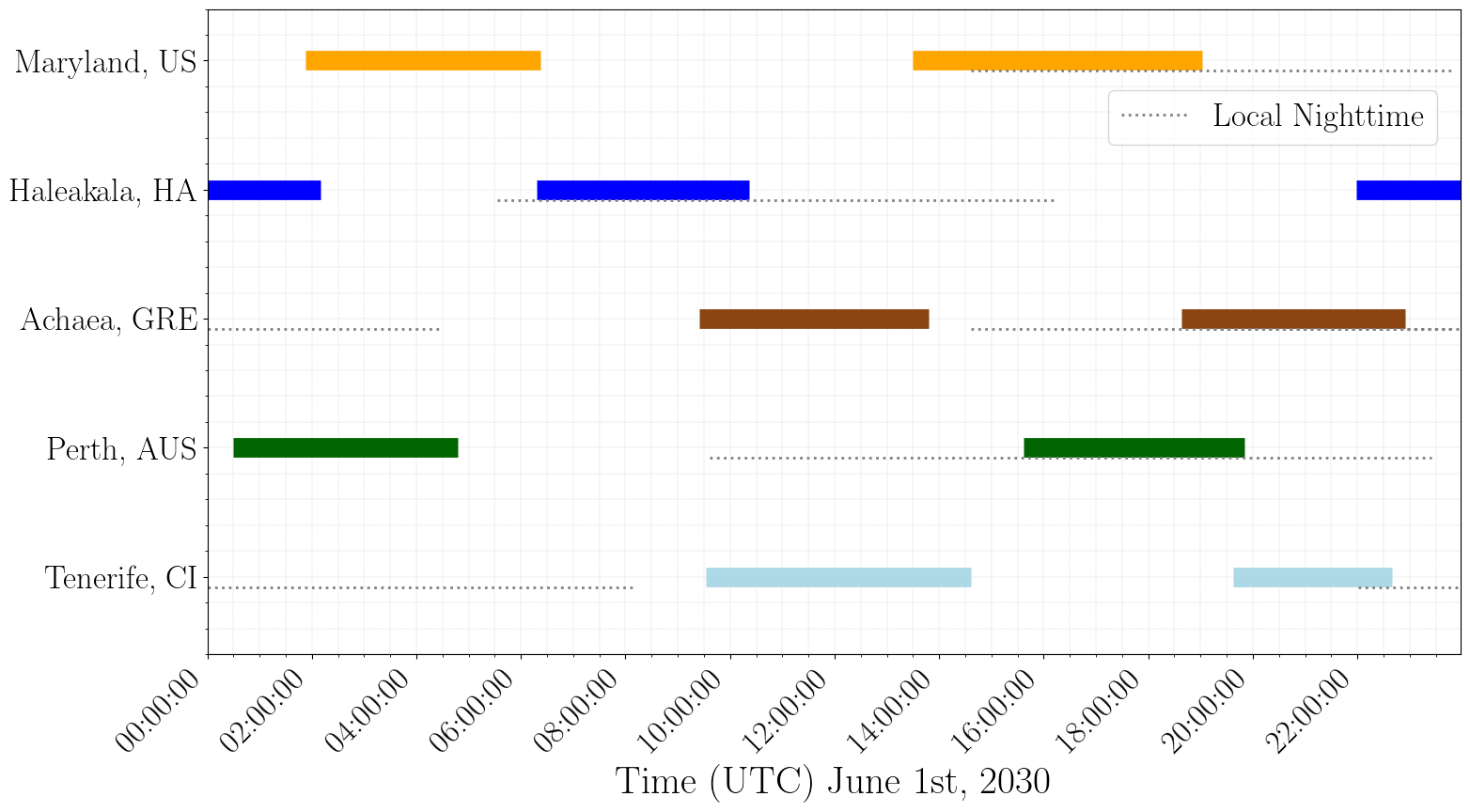} 
  \caption{Visibility of BHEX in a full day for a subset of downlink terminals in consideration for the laser communications network for (left) \m87 and (right) \sgra. Telescope elevations were calculated for each site, constrained to a maximum of 15 degrees due to telescope tilt limitations. The plots show the trajectory of BHEX, with the x-axis representing the 24-hour cycle and the y-axis indicating satellite coverage visible from Earth. A dotted line indicates local nighttime converted to a universal time zone.}
\label{fig:tracks}
\end{figure}

\begin{figure}[h]
    \centering
    \begin{minipage}{0.5\textwidth}
        \includegraphics[width=\textwidth]{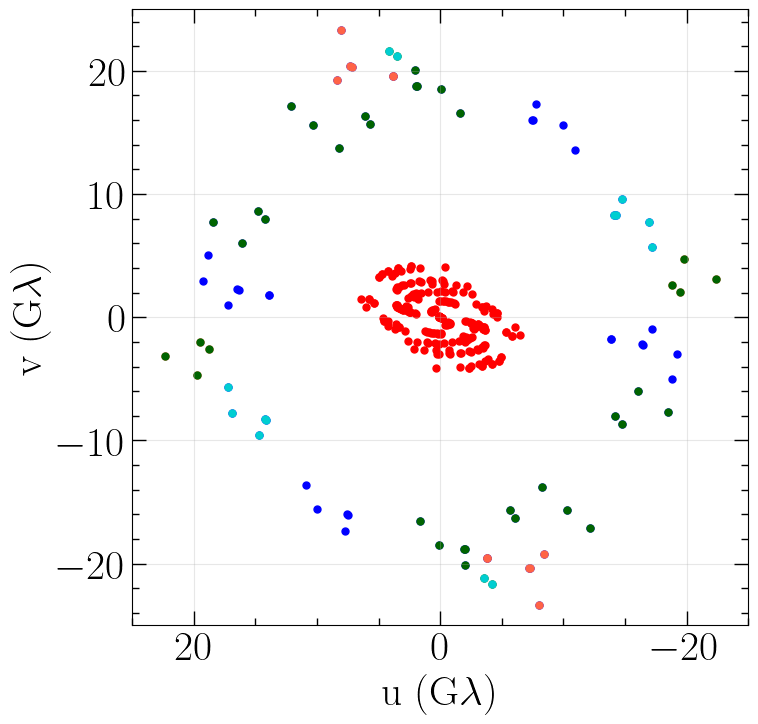}
    \end{minipage}\hfill
    \begin{minipage}{0.50\textwidth}
        \begin{subfigure}[b]{0.47\textwidth}
            \includegraphics[width=\textwidth,valign=c]{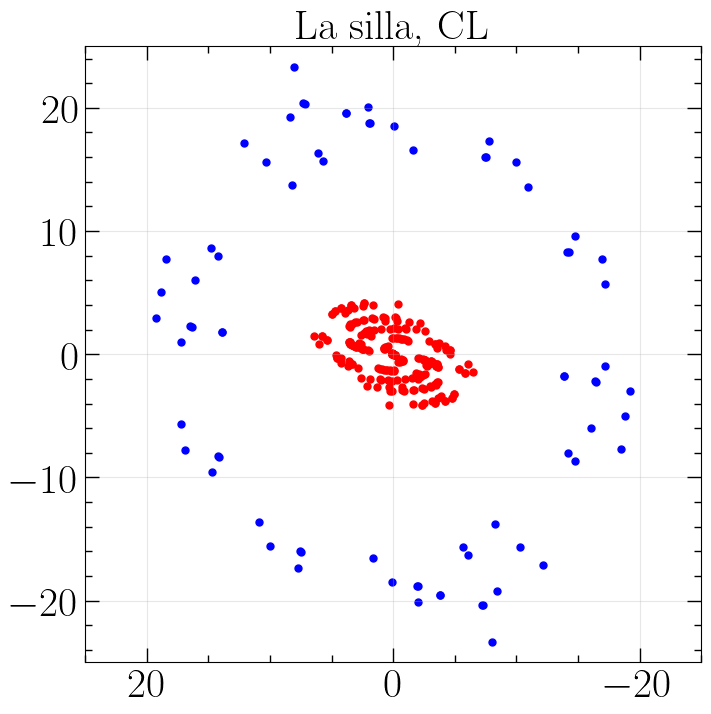}
        \end{subfigure}\hfill
        \begin{subfigure}[b]{0.47\textwidth}
            \includegraphics[width=\textwidth,valign=c]{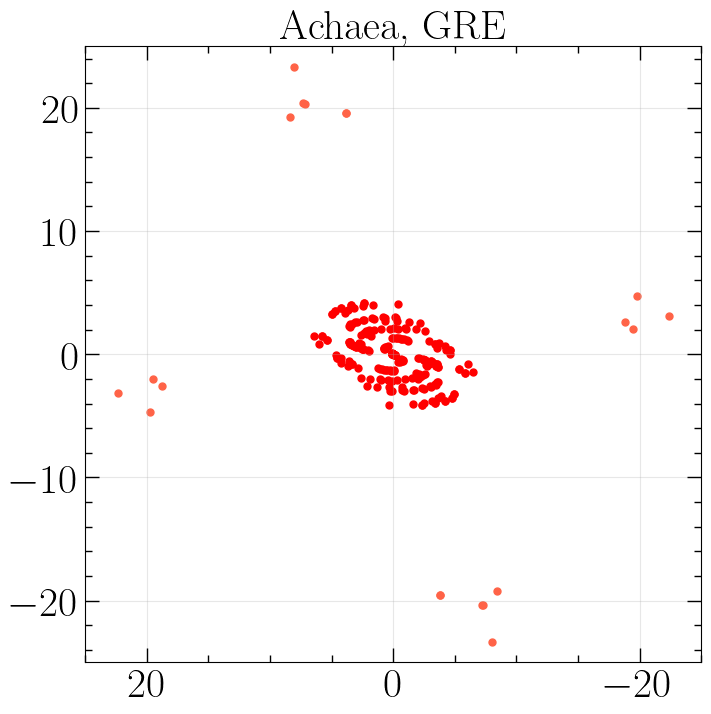}
        \end{subfigure}\\
        \begin{subfigure}[b]{0.47\textwidth}
            \includegraphics[width=\textwidth,valign=c]{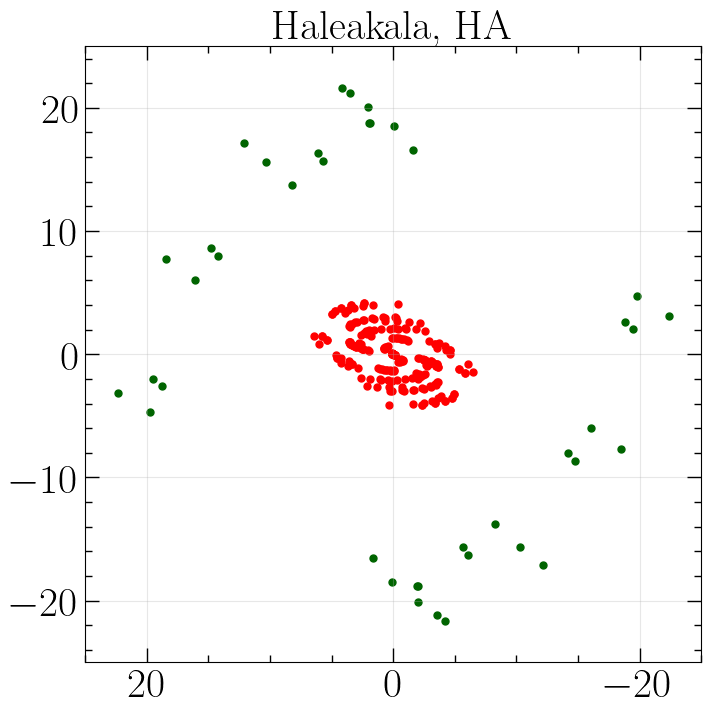}
        \end{subfigure}\hfill
        \begin{subfigure}[b]{0.47\textwidth}
            \includegraphics[width=\textwidth,valign=c]{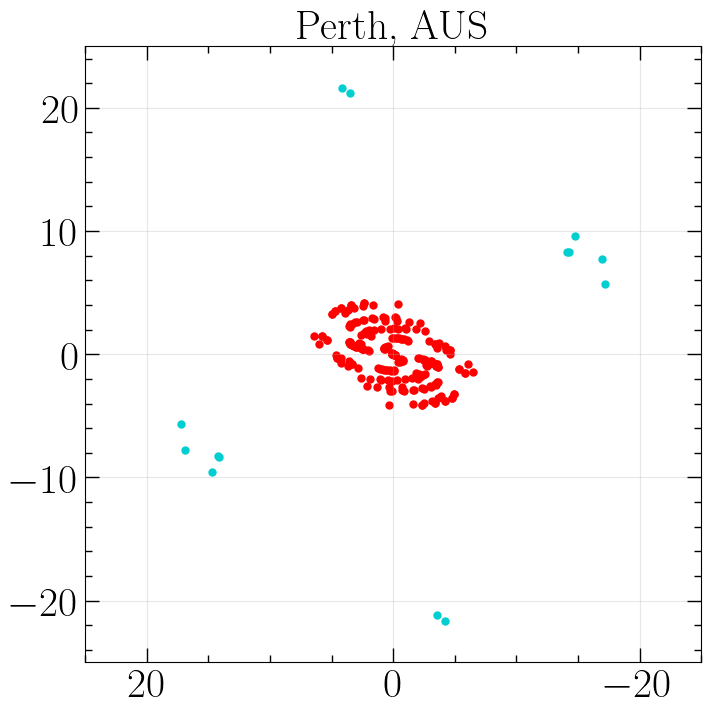}
        \end{subfigure}
    \end{minipage}
    \caption{An overview of simulated $(u,v)$ coverage for a single night of observations of \m87 with BHEX, colored by the portion of the BHEX data associated with each downlink terminal. Through simulated BHEX observations from a 12-hour circular orbit at medium Earth orbit, we visualize the radial filling of coverage as we sample increasing volumes of data. The plot on the left illustrates the combined coverage for all potential downlink stations, while the four plots on the right details the individual coverage in Chile, West Australia, Greece and Hawaii. The red points in the center are data points from ground-ground baselines.}
\label{fig:uvcoverage}
\end{figure}

In order to facilitate a 12-hour circular orbit in medium-Earth orbit, we down-selected four optical terminals among multiple locations. Reaching a balanced site distribution is essential to maintain complete coverage throughout the orbit. By covering a specific percentage of the orbit, each ground terminal will shorten the downlink path and lower the transmission latency. This is essential for obtaining real-time observations. A set of simulations were conducted to generate $(u,v)$ coverage expected for photon ring target observations, to show a representation of the predicted coverage pattern of the data downlink from BHEX. We take into account the elevation for each terminal and a 15-degree telescope tilt cut-off. In this setting, the selected stations covering a full 24-hour downlink were in La Silla, Chile, Perth, Australia, Haleakala, Hawaii, and Achaea, Greece. The orbit coverage of each terminal is shown in Figure~\ref{fig:M87_potential_stations}. For both \m87 and \sgra, a subset of downlink terminal sites were determined, most of which track the satellite during the local night, see Figure~\ref{fig:tracks}. The ESO telescope in Chile demonstrated the most extensive coverage, as shown by the $(u,v)$ plots in Figure~\ref{fig:uvcoverage}. The terminal was able to maintain high elevation angles for long periods of time due to its strategic location and favorable weather, which significantly reduced possibility of data loss. With minimal constraints, this shows that just one strategically positioned station that can cover substantial radial coverage on \m87. Further redundancy adds flexibility in case of weather/technical issues and observing more targets. 

We have carried out an extensive study detailing optical telescope options that fulfill at least a subset of criteria. As we explore changes to the satellite orbit, our requirements may vary and thus bring other sites into the forefront for an optical network partnership. Given the fast progress in space-based laser communication in the last few years\cite{BHEX_Wang_2024} and a growing interest in high-bandwidth data rates from space, we expect more facilities to be designed and operated specifically to service space mission optical downlink in the near future.

\section{Data Delivery}
Following the correlation model for the EHT, the BHEX data will be correlated with the ground array through existing infrastructure such as the MIT Haystack Observatory correlator and the Cannon Cluster at the Massachusetts Green High Performance Computing Center (MGHPCC). The correlated data will be processed using the EHT reference pipeline for basic quality checks, and then released to a public archive for science utilization. 

\section{Summary}
The BHEX concept of operations requires a three-part hybrid observatory with a space component, a ground science component, and a ground downlink network. Operating for a two-year mission, the photon ring science campaigns will take place in January-March for \m87 and in June-August for \sgra. Observations will be carried out with partner ground facilities, with participation of telescopes and time allocation varying with the requirements of the different science goals. Based on the current orbit, optical downlink is achievable continuously with a set of 4-5 downlink terminals. As requirements relax based on night-time observing and observing program cadence, only a subset of terminals could be needed at a given time. 

This unique mission concept is enabled by emerging space technologies such as laser communications and partnerships with the world's best radio observatories. BHEX will leverage billions of dollars of ground infrastructure for its operations, and add a space extension to an already established and well-exercised VLBI network that has delivered the transformative first images of black holes. BHEX will sharper these images to trace light at the edge of the universe, encoding properties of the black holes themselves, such as the spin that powers relativistic jets. The intersection of the successes of the EHT and demonstrations of laser communication in space make the path to the edge of a black hole within reach with BHEX.

\appendix

\acknowledgments 

This work is supported by the NASA Hubble Fellowship grant HST-HF2-51482.001-A awarded by the Space Telescope Science Institute, which is operated by the Association of Universities for Research in Astronomy, Inc., for NASA, under contract NAS5-26555. Technical and concept studies for BHEX have been supported by the Smithsonian Astrophysical Observatory, by the Internal Research and Development (IRAD) program at NASA Goddard Space Flight Center, by the University of Arizona, and by the ULVAC-Hayashi Seed Fund from the MIT-Japan Program at MIT International Science and Technology Initiatives (MISTI). We acknowledge financial support from the Brinson Foundation, the Gordon and Betty Moore Foundation (GBMF-10423), the National Science Foundation (AST-2307887, AST-2307888, AST-2107681, AST-1935980, and AST-2034306), the Simons Foundation (MP-SCMPS-00001470), and the Israel Science Foundation (grant \#2047/23). This project/publication is funded in part by the Gordon and Betty Moore Foundation (Grant \#8273.01). It was also made possible through the support of a grant from the John Templeton Foundation (Grant \#62286).  The opinions expressed in this publication are those of the author(s) and do not necessarily reflect the views of these Foundations. BHEX is funded in part by generous support from Mr. Michael Tuteur and Amy Tuteur, MD. BHEX is supported by initial funding from Fred Ehrsam. 

\bibliography{report} 
\bibliographystyle{spiebib} 

\end{document}